\documentclass[modern]{aastex701}
\NewPageAfterKeywords
\begin{document}

\title{An Orbit for a Massive Wolf-Rayet Binary in the LMC: \\ An Example of Binary Evolution\footnote{This paper includes data gathered with the 6.5 meter Magellan Telescopes located at Las Campanas Observatory, Chile. It also utilizes observations from the International Gemini Observatory, a program of NSF NOIRLab, which is managed by the Association of Universities for Research in Astronomy (AURA) under a cooperative agreement with the U.S. National Science Foundation (NSF) on behalf of the Gemini Observatory partnership: the NSF (United States), National Research Council (Canada), Agencia Nacional de Investigación y Desarrollo (Chile), Ministerio de Ciencia, Tecnología e Innovación (Argentina), Ministério da Ciência, Tecnologia, Inovações e Comunicações (Brazil), and Korea Astronomy and Space Science Institute (Republic of Korea).
}}

\author[0009-0006-4163-6402]{Breelyn Cocke}
\altaffiliation{2024-25 NAU/NASA Space Grant Student.}
\affiliation{Lowell Observatory, 1400 W Mars Hill Road, Flagstaff, AZ, 86001, USA}
\affiliation{Department of Astronomy and Planetary Science, Northern Arizona University (NAU), Flagstaff, AZ, 86011-6010, USA}
\email{bnc262@nau.edu}

\author[0000-0001-6563-7828]{Philip Massey}
\affiliation{Lowell Observatory, 1400 W Mars Hill Road, Flagstaff, AZ, 86001, USA}
\affiliation{Department of Astronomy and Planetary Science, Northern Arizona University (NAU), Flagstaff, AZ, 86011-6010, USA}
\email{phil.massey@lowell.edu}

\author[0000-0003-2535-3091]{Nidia I. Morrell}
\affiliation{Las Campanas Observatory, Carnegie Observatories, Casilla 601, La Serena, Chile}
\email{nmorrell@carnegiescience.edu}

\author[0000-0002-9109-5951]{Laura R. Penny}
\affiliation{Department of Physics and Astronomy, College of Charleston, 321 Rita Hollings Science Center, 58 Coming Street, Charleston, SC 29424, USA}
\email{PennyL@cofc.edu}

\author[0000-0002-5787-138X]{Kathryn F. Neugent}
\affiliation{Lowell Observatory, 1400 W Mars Hill Road, Flagstaff, AZ, 86001, USA}
\email{kathrynneugent@gmail.com}

\author[0000-0002-1722-6343]{Jan J. Eldridge}
\affiliation{Department of Physics, University of Auckland, Private Bag 92019, Auckland, New Zealand}
\email{j.eldridge@auckland.ac.nz}

\author[0000-0002-0548-8995]{Micha{\l} K. Szyma{\'n}ski}
\affiliation{Astronomical Observatory, University of Warsaw, Al. Ujazdowskie 4,
00-478 Warszawa, Poland}
\email{msz@astrouw.edu.pl}

\author[0000-0001-5207-5619]{Andrzej Udalski}
\affiliation{Astronomical Observatory, University of Warsaw, Al. Ujazdowskie 4,
00-478 Warszawa, Poland}
\email{udalski@astrouw.edu.pl}

\author[0009-0008-3389-9848]{Laurella C. Marin}
\altaffiliation{2024 REU student Lowell/NAU.}
\affiliation{Lowell Observatory, 1400 W Mars Hill Road, Flagstaff, AZ, 86001, USA}
\affiliation{Department of Astronomy and Planetary Science, Northern Arizona University (NAU), Flagstaff, AZ, 86011-6010, USA}
%\affiliation{Department of Astronomy, Box 351580, University of Washington, Seattle, WA 98195, USA}
\email{laurella.c.marin.25@dartmouth.edu}

\begin{abstract}
Wolf-Rayet (WR) stars are helium-burning, evolved massive stars which
have had most of their hydrogen-rich outer layers removed either
through stellar winds and/or binary stripping. Here we report on
LMC173-1, a WN3+O binary located in the Large Magellanic Cloud
(LMC).  Using spectra obtained from the Magellan and Gemini-S
telescopes, we have derived system parameters for this intriguing
binary.  The WR star's mass is only 43\% that of its companion, and we argue that this requires binary evolution rather
than mass loss by stellar winds alone, given the metallicity of the
LMC.  The stars are close enough to each other with their 3.52 day period
that the O star is actually orbiting within the wind of the WR star,
as is the case for other well-known WR+O systems, such as V444 Cyg.  As
a result, high precision OGLE photometry reveals a WR atmospheric
eclipse,  as well as a 7-8
millimag ellipsoidal modulation due primarily to the tidal distortion of the O star.  
Modeling the light curve allows us to estimate the orbital inclination.
Derivation of stellar parameters suggests neither component is filling
its Roche-lobe surface today. The O star is spinning much faster than synchronous rotation.
Using BPASS v2.2 binary models, we discuss the
probable evolutionary history of the system.  The WR progenitor 
likely underwent Case A Roche-lobe overflow (RLOF) before leaving the
main-sequence. As it lost its H-rich envelope, it became a WN-type
WR. The resulting system is a binary with similar luminosities but
very different radii, representing a post-RLOF phase.

\end{abstract}

\section{Introduction}

Massive stars influence a wide range of astrophysical phenomena. They are the precursors of gamma ray bursts, black holes, and neutron stars (see, e.g., \citealt{1993ApJ...405..273W}). They contribute to the chemical enrichment of the Universe, including the elements necessary for life (see, e.g., \citealt{JJohnson}). Binary massive stars may evolve to binary black holes or neutron stars, the merger of which generates gravitational waves, disturbing the curvature of space (see, e.g., \citealt{2013ApJ...779...72D,2017ApJ...846..170T}). 

A subset of massive stars, the Wolf-Rayet (WR) stars, are the helium burning descendants of massive O-type stars. They are in their last stage of evolution before undergoing core collapse. WR stars are characterized by their unique spectra. They present broad, strong emission lines of He and N (classified as WN-type stars) or C and O (classified as WC-type stars). Most WRs contain little or no hydrogen. Rather, their surfaces show either the products of the H-burning CNO cycle (WNs) or triple-$\alpha$ He-burning (WCs). This suggests that the outer hydrogen-rich layers have been partially or completely removed, exposing the material processed in the inner layers.  This can happen either through binary interactions \citep{1967AcA....17..355P}, and/or through the actions of the progenitor's strong stellar winds while on the main sequence \citep{1975MSRSL...9..193C}.

An excellent candidate to study this process is the star [MNM2014] LMC173-1. 
The star was identified as a WR candidate during the first observing run of a four-year on-band, off-band interference-filter imaging survey for WRs that eventually covered all of LMC and SMC galaxies \citep{FinalCensus}.  An optical spectrum taken in 2013 confirmed that the star was indeed a previously unknown WR.  Based upon that spectrum, \citet{MasseyMCWRI} classified the star as WN3 + O7~V.  Over the years we continued to observe the star as opportunity arose in the hopes of obtaining an orbital solution, and hence information about its physical properties.  We find here that the WN3 component has a mass only 43\% of its O-type companion, and yet as the more evolved object, it must have started out as the more massive object. That would make it hard to achieve via stellar winds alone at the LMC's metallicity, making this an interesting object to study from the perspective of binary evolution.

In Section~\ref{Sec-obs}, we describe our observations and reductions using the Magellan and Gemini telescopes and instruments. In Section~\ref{Sec-analysis}, we describe what we learn from the spectra, starting with the physical parameters, and proceeding to the measurement of the radial velocities and determining the orbit solution.  In Section~\ref{Sec-lc} we give our analysis of the light-curve, and what this means in terms of the system parameters.  Finally in Section~\ref{Sec-discussion}, we present a plausible evolutionary history of the system.

\section{Observations and Reductions}
\label{Sec-obs}

\subsection{Magellan}
Our initial spectroscopy was carried out using the Magellan Echelette (MagE, \citealt{MagE}), which provided coverage from 3200~\AA\ to 1$\mu$m at a spectral resolving power of 4100 (73 km s$^{-1}$) using a 1\arcsec\ slit. The detector is a single E2V 42-20 chip with a 2048$\times$1024 format with 13.5 $\mu$m pixels. The plate scale is 0\farcs3 pixel$^{-1}$. For the 2013 observations the instrument was mounted on one of the auxiliary ports of the Clay 6.5 meter Magellan telescope; subsequently, it was moved to the Baade 6.5 meter.  

The spectra were taken as a secondary goal during time assigned to other projects.  Exposure times varied from 450~s to 900~s depending upon the seeing and circumstances.  A 3 s Th-Ar was made after each observation prior to moving the telescope in order to provide accurate wavelength calibration. The observations were all made by N.I.M., P.M., and/or K.F.N., except for the final MagE spectrum, which was kindly obtained by Dr.\ Marcelo Mora. 

Bias frames were typically run each afternoon, followed by flat-field exposures, which were used primarily to remove fringing in the red.
Spectrophotometric standards were observed each night in order to provide fluxing needed for the merging of the different echelette orders. 

Reductions were carried out using a combination of {\sc mtools} routines (written by Jack Baldwin) and {\sc iraf} echelle tasks, as described in \citet{2012ApJ...748...96M}.

In all, we obtained 13 spectra with MagE. The observations are listed in Table~\ref{tab:RVs}.

\begin{deluxetable}{l c l r r r}
\tablecaption{\label{tab:RVs} Radial Velocities of LMC173-1}
\tablehead{
\colhead{HJD}
&\colhead{UT Date}
&\colhead{Tel/Inst}
&\colhead{Phase\tablenotemark{a}}
&\multicolumn{2}{c}{RV (km s$^{-1}$)} \\ \cline{5-6}
& & & &  O Star\tablenotemark{b}&WR\tablenotemark{c} 
}
\startdata
2456581.700 & 2013 Oct 16 & Clay/MagE     &  0.95  &  266.7 &    -0.1\\
2456640.694 & 2013 Dec 14 & Clay/MagE     &  0.71  &  370.4 &  -181.2\\
2458118.705 & 2017 Dec 31 & Baade/MagE    &  0.46  &  213.3 &   114.7\\
2458124.814 & 2018 Jan 06 & Baade/MagE   &  0.20  &  157.7 &   322.0\\
2458153.785 & 2018 Feb 04 & Baade/MagE    &  0.43  &  199.4 &   227.9\\
2459192.667 & 2020 Dec 09 & Baade/MagE    &  0.47  &  222.1 &   164.1 \\
2459236.786 & 2021 Jan 22 & Baade/MagE    &  0.00  &  234.4 &    97.3\\
2459237.787 & 2021 Jan 23 & Baade/MagE    &  0.28  &  154.8 &   330.9\\
2459568.807 & 2021 Dec 20 & Baade/MagE    &  0.29 &  159.0 &   272.9 \\
2459569.759 & 2021 Dec 21 & Baade/MagE    &  0.56  &  286.6 &   -71.9\\
2459854.864 & 2022 Oct 02 & Baade/MagE    &  0.53  &  256.0 &   -30.0\\
2459883.779 & 2022 Oct 31 & Baade/MagE    &  0.74  &  374.9 &  -216.4\\
2460570.898 & 2024 Sep 17 & Baade/MagE    &  0.89  &  325.8&  -139.9 \\
2460672.739 & 2024 Dec 28 & Gemini-S/GMOS &  0.81  &  356.6 &  -203.1\\
2460674.725 & 2024 Dec 30 & Gemini-S/GMOS &  0.37 &   156.8   & 192.8\\
2460680.554 & 2025 Jan 05 & Gemini-S/GMOS &  0.03  &  208.6 &    53.9\\
2460706.639 & 2025 Jan 31 & Gemini-S/GMOS &  0.44  &  208.4 &   112.2 \\
2460712.543 & 2025 Feb 06 & Gemini-S/GMOS &  0.11  &  190.1 &   173.1\\
2460713.550 & 2025 Feb 07 & Gemini-S/GMOS &  0.40  &  196.2 &   172.7\\
\enddata
\tablenotetext{a}{Computed using P=3.521120 d and phase zero-point T0=2456588.91 (corresponding to the WR star being in inferior conjunction) as derived in Section~\ref{Sec-orbit}.
}
\tablenotetext{b}{Average (median) of 7 absorption lines. Typical standard deviation of the mean is 6.9 km~s$^{-1}$.}
\tablenotetext{c}{Zero-point of the RV of the WR star was set by assigning the laboratory wavelength of the N\,{\sc v} $\lambda\lambda$ 4604,20 doublet to 4611.92~\AA, roughly the average of the lab values for the two components. Typical measuring uncertainty is  41 km~s$^{-1}$.}

\end{deluxetable}

\subsection{Gemini}

Despite the large number of MagE spectra, we could not coax a period determination out of the measurements.  We realized that one reason that the period had not been well constrained by the observations is that they had been taken scattered throughout 2013-2024. (See Table~\ref{tab:RVs}.) Given the large radial velocity variations we found, and the two occasions where two spectra were taken just one night apart, we believed the period was short, on the order of a few days. (That would also be consistent with significant binary mass loss.) We therefore applied for Gemini ``Fast-turnaround" time, a program by which a few observing hours might be allocated and the observations carried out during a 3 month period directly following.  

Our program was selected (GS-2024B-FT-212), and new spectra were taken in queue mode with the Gemini Multi-Object Spectrograph  on Gemini-South (GMOS-S). The detector package for GMOS-S consists of three 2048$\times$4176 Hamamatsu CCDs arranged in a row with $\sim$60 pixel gaps between adjacent chips. The data is read out through 12 amplifiers, i.e., one amplifier at each chip corner. The spatial scale is 0\farcs08 per unbinned pixel; we binned the chip to 1 (spectral) $\times$ 2 (spatial).  Since our target was a single point source being observed with a long slit, we restricted the readout to the central 80\arcsec\ region of interest within the 5\farcm5 long slit; this is the recommended setup to reduce readout times.

We used the B1200 grating centered at 4550~\AA, providing wavelength coverage from 3750~\AA\ to 5350~\AA, with two 16-18~\AA\ gaps due to the spacings between the three CCDs. The grating tilt was carefully chosen both to assure coverage down to the upper Balmer absorption lines, and to make sure that no spectral features of interest would fall into the gaps. The spectra were taken in long-slit mode with a 1\arcsec\ slit width.  This setup provided 0.26~\AA\ pixel$^{-1}$ dispersion, and a resolution of about 8 pixels (i.e., 2.1~\AA, or 160 km s$^{-1}$).\footnote{Although significantly lower in spectral resolution than the MagE observations, the resolution is similar to the line widths found in O-type dwarfs, and was quite adequate.} 
Each observation consisted of 3$\times$600 s exposures followed by a 12~s internal flat  and a 60~s CuAr arc exposure for wavelength calibration. 

The reductions were carried with the V1.16.1 Gemini external package for the modern V2.18 {\sc iraf} release that runs on Apple Silicon Mac computers. (See, e.g., \citealt{1986SPIE..627..733T,1993ASPC...52..173T}, and \citealt{2024arXiv240101982F}.) We followed a procedure guided by the V2 of the GMOS Data Reduction Cookbook.\footnote{https://nsf-noirlab.gitlab.io/csdc/usngo/gmos-cookbook/\#iraf-proc-ls} One problem we encountered was the use of ``gsskysub" to remove sky prior to extracting the spectrum.  At this point, the data have been ``straightened" (transformed) so that each column is at a constant wavelength.  However, this creates a problem if there are any bad columns greatly expanding their effect on the data.  A good workaround was to do the sky subtraction as part of the regular extraction process as done in standard long-slit {\sc iraf} reductions for other instruments.\footnote{https://noirlab.edu/science/sites/default/files/media/archives/documents/scidoc0479.pdf}  Throughout this process we received useful help from both Mike Fitzpatrick and Dr.\ Vinicius Placco at NOIRLab. As with the Magellan spectra, we then normalized the spectra before our analysis. 

We successfully obtained six spectra taken between 2024 Dec 28 and 2025 Feb 7.  We list these in Table~\ref{tab:RVs}. These provided a much higher frequency of coverage than the MagE data, and proved to be  just what we needed in order to find the period and provide complete phase coverage.

\section{Analysis}
\label{Sec-analysis}

In this section we will first discuss the stellar parameters (temperatures, flux ratio, luminosities, and radii), and then proceed to discuss the measurement of the radial velocities, determination of the period, and orbital calculation.   In discussing the spectra we refer to the orbital phases even though those will not be derived until the following subsection.

We normalized all 19 of our spectra by fitting a low order cubic spline to the data. The Magellan spectra had been flux calibrated in order to combine the echelette orders.  The Gemini data were obtained using a first-order grating, and so flux calibrating was not needed; nor was it desirable as we would normally normalize the data before attempting any analysis, particularly the measurement of radial velocities. In order to facilitate the normalization, we first cropped the Magellan data to the 3700-5000~\AA\ region to better match the Gemini coverage and to reduce the demands on the fitting function over a larger wavelength region.  The only interesting spectral features this excluded were He\,{\sc i} $\lambda$5876 and the H$\alpha$/He\,{\sc ii} blend, but for those we had the fluxed versions available for inspection.

\subsection{The Spectrum and Derived Physical Parameters}

\begin{figure}
% \epsscale{0.9}
\plotone{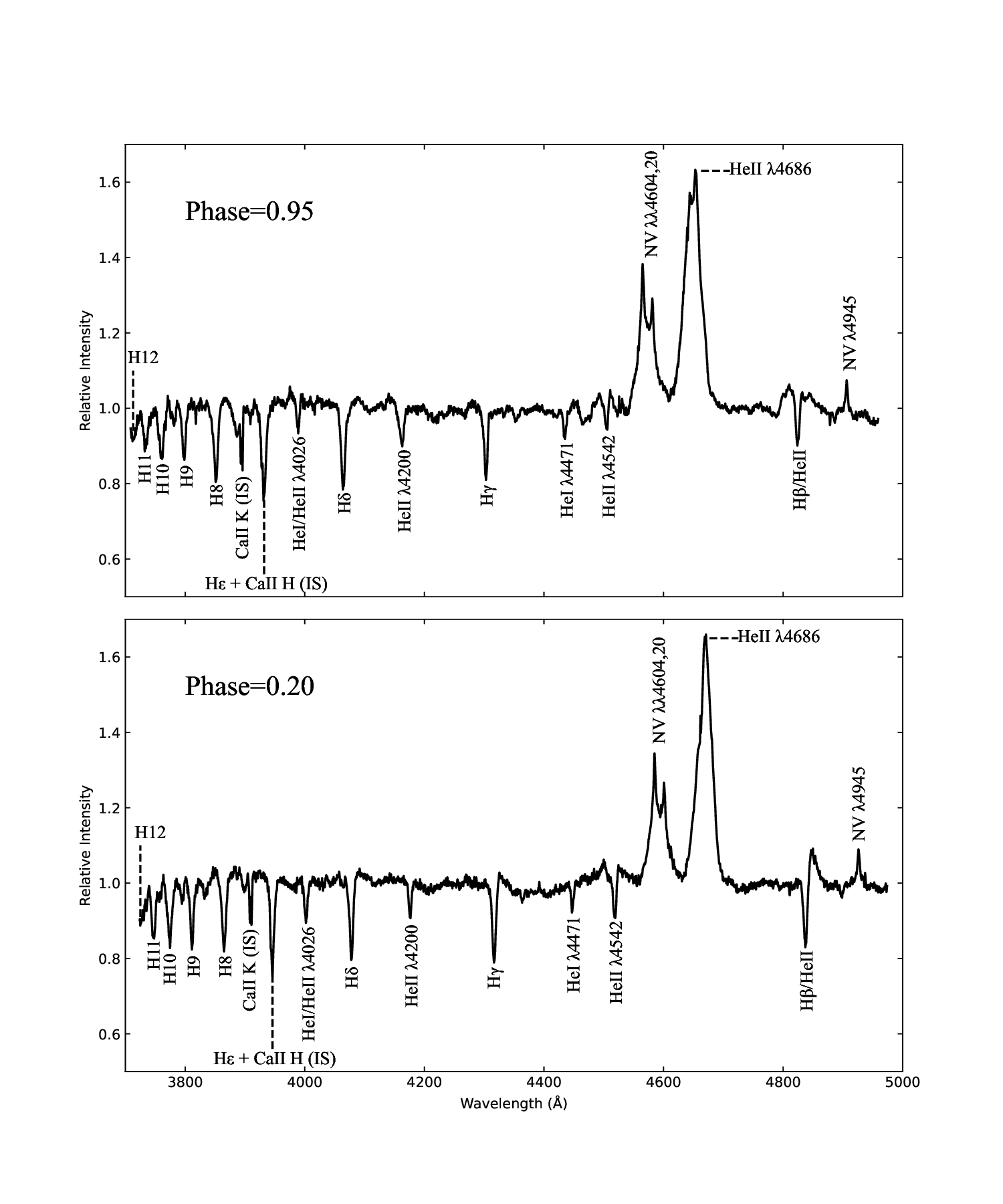}
\caption{\label{fig:spect}Blue spectral region of LMC173-1.  The upper spectrum was taken on 2013 Oct 16 at a phase of 0.95, and the lower spectrum on 2016 Jan 6, at a phase of 0.20.  Note the difference in the line strength of He\,{\sc ii} $\lambda$4542, due to the relative shifting of weak broad WR emission with the  narrower absorption from the O star. The prominent lines are labeled.}
\end{figure}

As mentioned above, \citet{MasseyMCWRI} classified the star as WN3 + O7~V.  The absorption line spectrum clearly dominates the combined spectrum (given their strength), as shown in Figure~\ref{fig:spect}.
The upper Balmer lines are strongly in absorption, and He\,{\sc i} and He\,{\sc ii} are both found in absorption.  The predominant WR emission features are N\,{\sc iv} $\lambda$3480, the strong N\,{\sc v} $\lambda \lambda$4604,20 doublet, and He\,{\sc ii} $\lambda$4686.  The lack of N\,{\sc iv} $\lambda$4058 or N\,{\sc iii} $\lambda \lambda$4634,42 results in a WN3 classification for the WR.  At longer wavelengths (e.g., H$\beta$ and H$\alpha$, the later of which is not shown in the figure), the
corresponding WR He\,{\sc ii} line emission begins to dominate over the O-star absorption, serving as a reminder that the other hydrogen Balmer and He\,{\sc ii} absorption lines may be affected by emission as well as continuum dilution. The phase-dependent velocity shift between the WR and O stars  leads to variability in the strengths of the He\,{\sc ii} and lower Balmer lines, as illustrated in
Figure~\ref{fig:spect}. The luminosity class of the O-type star is impossible to determine on the basis of the usual spectroscopic criteria of He\,{\sc ii} $\lambda$4686 absorption, as this is swamped by the WR's emission, but like
 \citet{MasseyMCWRI}, we argue that it must be a dwarf, based on the 
 composite $V$ magnitude of $\sim$14.65, corresponding to $M_V\sim-4.25$.

 \subsubsection{Effective Temperature of the O Star}
\begin{figure}
 \epsscale{0.49}
 \plotone{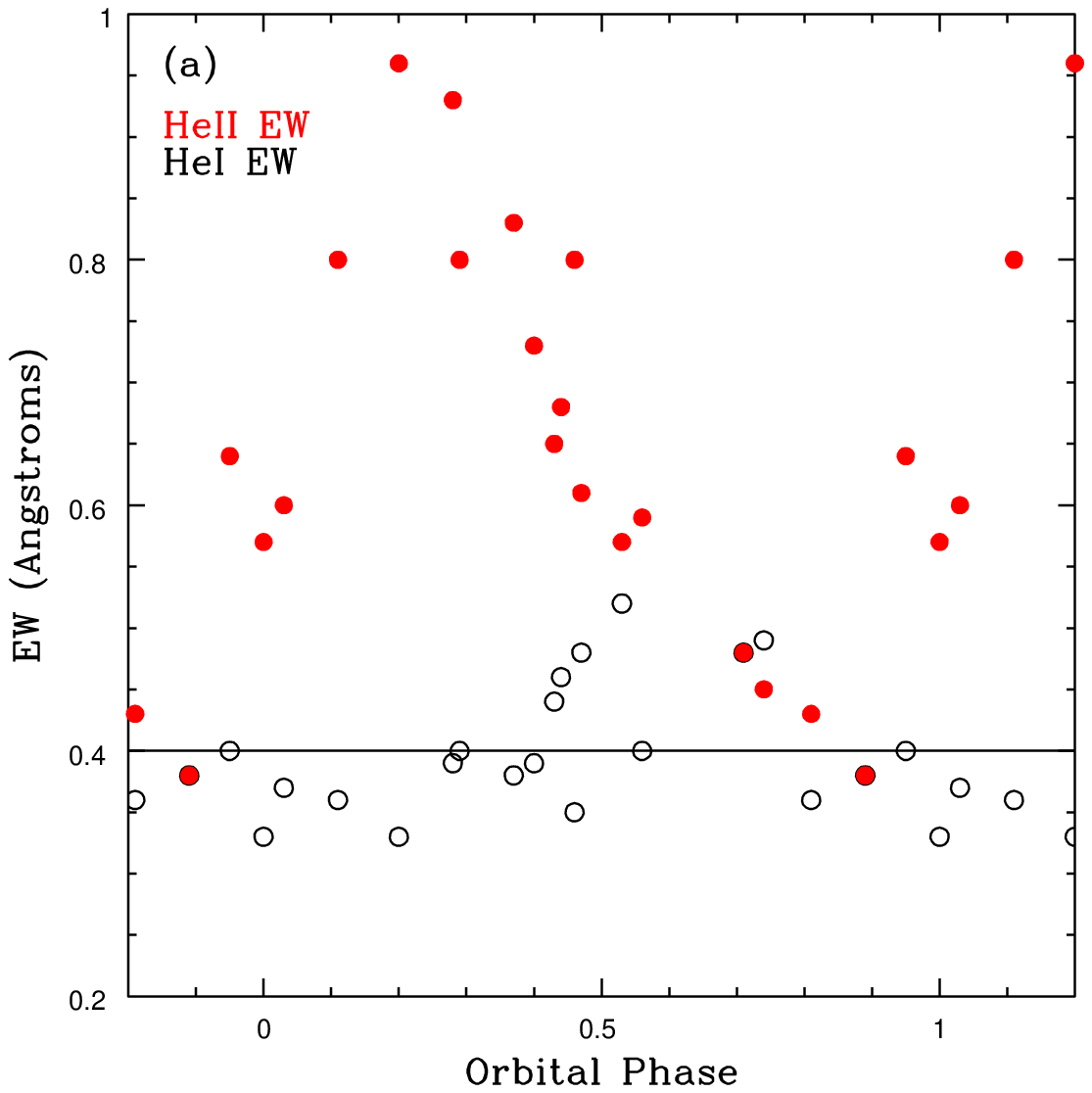}
 \plotone{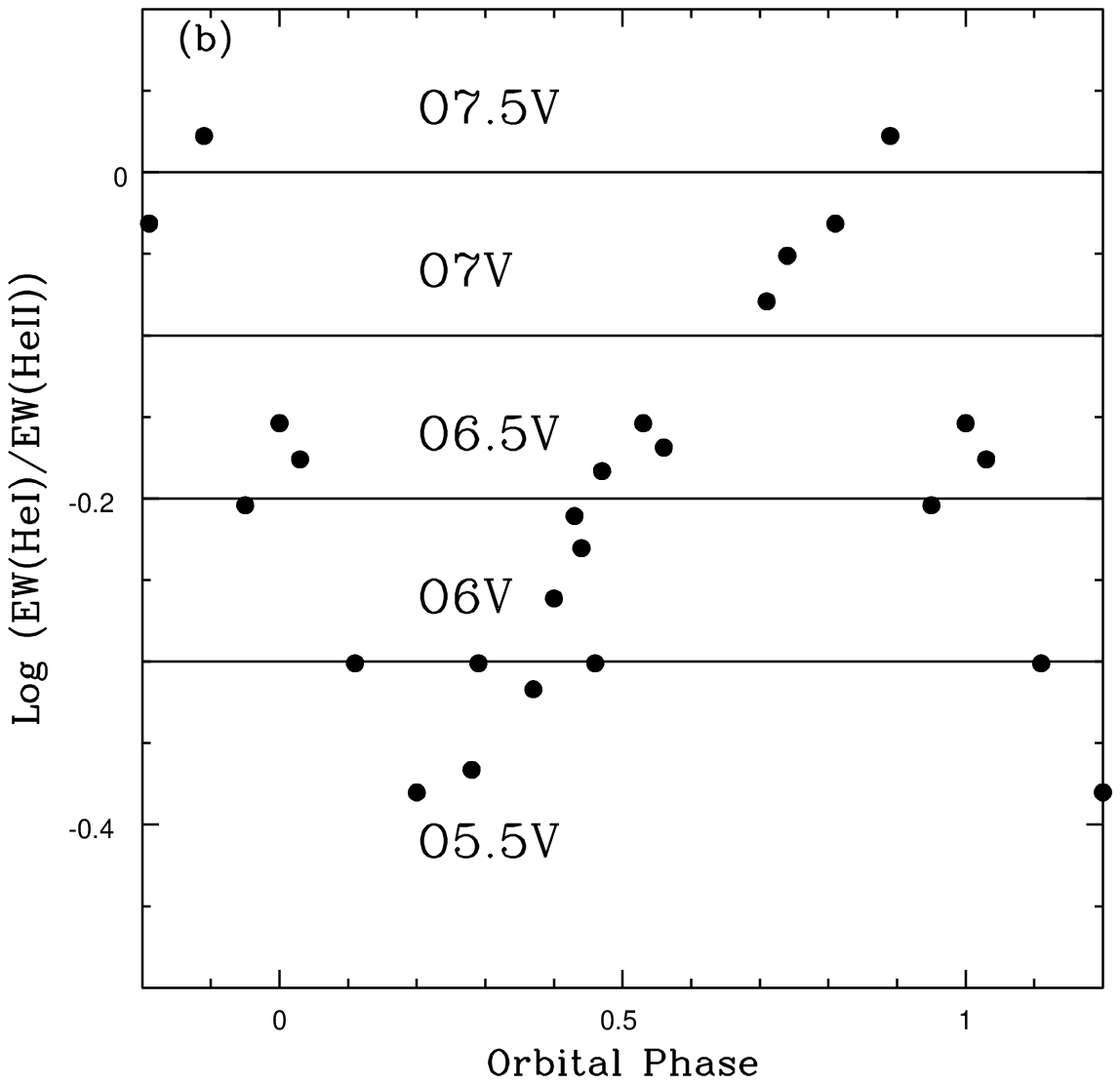}
 \caption{\label{fig:EWs} EW variations with orbital phase.  (a) The EWs of He\,{\sc i} $\lambda$4471 are shown by black open circles, and those of He\,{\sc ii} $\lambda$4542 are shown by filled red circles.
 The line shows the median He\,{\sc i} EW value of 0.40~\AA. (b) The ratios of the EWs were used by \citet{1973ApJ...179..181C} to define the O-type spectral classes, with the ranges indicated in the figure.  We adopted the median He\,{\sc i} EW value in these calculations.  The star's spectral type varies from O5.5 to O7.5 during its cycle,  but the constancy of the He\,{\sc i} EW shows that the temperature itself is not changing.}
 \end{figure}

 The spectral subtype of O stars is determined primarily by the He\,{\sc i} $\lambda$4471 to He\,{\sc ii} $\lambda$4542  ratio (see, e.g., \citealt{2009ssc..book.....G}).  \citet{MasseyMCWRI} show a spectrum with the He\,{\sc i} line only slightly weaker than the He\,{\sc ii} line, leading to the O7~V classification.  To better determine the star's temperature, we measured the equivalent width  (EW) values in order to use the quantitative classification scheme defined by \citet{1973ApJ...179..181C}.  We quickly found that while the EW of He\,{\sc i} $\lambda$4471 was consistently measured as $0.40\pm0.02$~\AA, the EW of the He\,{\sc ii} $\lambda$4542 strongly varied with phase (see Figure~\ref{fig:spect}).  It was clear that faint He\,{\sc ii} emission from the WR star was contaminating the absorption line measurements, and that as the WR star moved one way in its orbit and the O star in the other, the amount of contamination came and went.  The strongest He\,{\sc ii} absorption is at a phase near 0.25.  In Figure~\ref{fig:EWs} (left) we show how the EWs of the He\,{\sc i} $\lambda$4471 and He\,{\sc ii} $\lambda$4542 vary with orbital phase.  On the right, we show how the log of the ratio of the EWs varies, indicating what this means for the spectral type.
 
 We emphasize that we are not suggesting that the effective temperature of the O star is changing with orientation.  If that were the case we would see a strong variation in the EW of the He\,{\sc i} line, as its strength is quite sensitive to temperature in this regime.   Instead, we are simply seeing the effect of a shifting underlying He\,{\sc ii} line emission on top of the absorption feature. We therefore suggest that the earliest spectral type, O5.5~V, is a far better representation of the temperature than the original ``O7~V" type given by \citet{MasseyMCWRI}. (Note that their spectral type was based on their 2013 spectra taken
 at phases of 0.95 and 0.71.)   Using their Galactic-metallicity {\sc cmfgen} calibration of O-type spectral types, \citet{2005A&A...436.1049M} assign a temperature of 40,000 to the O5.5~V spectral subtype. \citet{2013ApJ...768....6M} analyzed the 
 spectra of the LMC O5.5~V((f)) star Sk -70$^\circ$69 with {\sc fastwind} and {\sc cmfgen} finding effective temperatures of 39,500 and 41,000, respectively.  We therefore conclude that 40,000 is a good  value of the temperature of the O star companion. Technically this is a lower limit, as there may still be some contamination of the He\,{\sc ii} line even at its strongest.  However, given the EW of He\,{\sc i} $\lambda$4471, the effective temperature cannot be much higher, as we discuss below when we determine the flux ratio between the two stars. 
 
 \subsubsection{Effective Temperature of the WN3 Star}
 
The ``effective temperature" of the WN3 star is even harder to determine.  WRs have very extended atmospheres; all of the spectral lines and continua are formed in an expanding, outflowing wind. For our purposes here, we will use the temperature at an optical depth of 2/3rds, a good representation from whence the continuum radiation originates.

The spectral subtype of WN3s is defined by the presence of N\,{\sc v} $\lambda \lambda$4604,20 and the lack of N\,{\sc iv} $\lambda$4058 \citep{vanderHuchtII}. Although there is an even higher excitation class, WN2, with no N\,{\sc v}, the WN3 subtype is a bit of a catchall; stars of a large range of excitation temperatures can fall into this subtype (see, e.g., \citealt{PotsLMC}).  Co-author P.M. is leading a study using {\sc cmfgen} to model new data on LMC WN3-4 stars.  This study has not yet been completed, but for the six WN3 stars with advanced models,  the temperatures at an optical depth of 2/3rds span a range from 49,200 to 73,000~K.  The average value of these six stars is 60,650~K, and the median value is 61,000~K.  We therefore adopt a value of 61,000$\pm$12,000~K for the WN3 component.
 
\subsubsection{The Flux Ratio}
In order to determine the flux ratio between the two components, $F_{\rm WR}/F_{\rm Ostar}$, we compare the EWs of several absorption lines to what we expect the intrinsic EWs would be if there was no companion present. If the companion star (the WR in this case) provides only continuum radiation at the wavelengths of the lines we use, then $$\frac{\rm EW_{\rm intrinsic}}{\rm EW_{\rm measured}} = 1 + \frac{F_{\rm WR}}{F_{\rm Ostar}},$$ and hence $$\frac{F_{\rm WR}}{F_{\rm Ostar}} = \frac{\rm EW_{\rm intrinsic}}{\rm EW_{\rm measured}}-1.$$

For the values of the intrinsic EWs, we constructed {\sc cmfgen} models for $\log g=4.0$, and $T_{\rm eff}$ of 39,000, 40,000, and 41,000.  To do so, we modified the Galactic metallicity ``NT40000\_logg400" model supplied with the software, and described by \citet{2005A&A...436.1049M}, to 1/2 solar to match the LMC metallicity ($z=0.007$).   Only the elements carbon, nitrogen, oxygen, silicon, phosphorus, sulfur, and iron were included.  The standard H/He number ratio of 10.0 was assumed. A turbulent velocity (``VTURB") of 20 km~s$^{-1}$ was assumed, and the output spectrum modified to a microturbulent velocity of 15 km s$^{-1}$. Wind parameters are not particularly important for mid-O type dwarfs; we included a small mass-loss rate ($<1\times10^{-6}$ M$_\odot$ yr$^{-1}$) with the standard $\beta=0.8$ acceleration parameter and a terminal velocity of 3000 km s$^{-1}$, roughly 2$\times$ the escape velocity.

The difficulty is that basically all but the He\,{\i} lines are contaminated at some level by He\,{\sc ii} emission from
the WR star.  The contamination is lesser at shorter wavelengths, and we found that H$\delta$ was the best line to use, in that measurements of its EW amongst the 19 spectra showed the smallest scatter.  At shorter wavelengths, the upper Balmer lines are simply too weak for reliable EW measurements, and at longer wavelengths, the contamination by the WR He\,{\sc ii} lines becomes prominent; see, e.g., H$\beta$ in Figure~\ref{fig:spect}.  For the 40,000~K model we obtained a flux ratio of 0.26 based on H$\delta$.
The He\,{\sc i} $\lambda$4471 yielded a value of 0.23 with this model, and the He\,{\sc i} $\lambda$5876 line a value of 0.30.  The 39,000~K model yielded the same result for H$\delta$ but a larger flux ratio for the He\,{\sc i} lines, while the 41,000~K model yielded the same result for H$\delta$ but a much lower flux ratio for the He\,{\sc i} lines; we take this as additional confirmation that the 40,000~K model represents the best value for the temperature of the O star, and we adopt a flux ratio of 0.25$\pm0.15$.  The uncertainty is an estimate based upon spectrum to spectrum variations in the measured EWs, and is intended as an outer bound. 

\subsubsection{Derived Physical Parameters and Uncertainties}

We adopt a distance of the LMC of 50.0~kpc (e.g., \citealt{vandenbergh2000} and references therein), corresponding to a true distance modulus of 18.49.\footnote{Note that this commonly accepted value for the distance to the LMC is in good accord with recent ``precise" measurements, such as \citet{2019MNRAS.490.4254N}, \citet{2019Natur.567..200P}, and \citet{2026arXiv260215421C}.} As for the reddening, \citet{Massey2002} measured  $B-V=-0.14$ for LMC173-1.  (We have used our spectrophotometry to determine that the effect of the WN's emission lines on this value is negligible.) Assuming the O star dominates the flux, the expected $(B-V)_0$ is $-0.27$ \citep{2005A&A...436.1049M}, leading to an $E(B-V)=0.13$.  This is identical to the typical $E(B-V)$ value measured for early-type stars throughout the LMC \citep{LGGSII}.   We assume that the uncertainty on this value is 0.05~mag.  This leads to an apparent distance modulus of 18.89$\pm$0.16.  The OGLE-IV photometry presented in Section~\ref{Sec-lc} gives $V=14.65$, with a negligible uncertainty.   This leads to an absolute visual magnitude of the system of $M_V=-4.24\pm0.16.$

A large source of uncertainty is in the visual flux ratio, $F_{\rm WR}/F_{\rm O star}=0.25\pm0.15$.  Again, we consider this uncertainty a plausible bound.  Using this, we derive $$M_V (\rm O star) = -4.01^{+0.28}_{-0.30}$$ and 
$$M_V (\rm WR star) =-2.50^{+0.42}_{-0.39}.$$ The individual absolute visual magnitudes were obtained by splitting the total absolute visual magnitude according to the observed visible flux ratio; plausible bounds were derived by jointly varying the total magnitude (from the extinction uncertainty) and the flux ratio within their adopted limits, with the extrema taken as the uncertainty ranges.

To convert to luminosity, we adopt the relationship between the effective temperature of hot OB stars and bolometric corrections determined by \citet{2005A&A...436.1049M}.  (Although this may not strictly apply to WR stars, the uncertainty introduced by using this approximation is doubtless less than that introduced by the large range of possible temperatures for the WR star.)  For the O star we adopt an effective temperature of 40,000$\pm500$, and for the WR star, an effective temperature of 61,000$\pm$12,000~K, as discussed above.  This leads to luminosities of $$\log L/L_\odot \rm{(O star)} =4.99\pm0.07,$$ and $$\log L/L_\odot \rm{(WR star)} =4.88^{+0.39}_{-0.58}.$$  Again, the uncertainties in the luminosities were determined by jointly varying the extinction-dependent absolute visual magnitude, the visual flux ratio, and the effective temperatures within their plausible bounds, with the luminosities of the two components constrained by the fixed total light; the extrema of the resulting distributions were adopted as the quoted ranges.

Next we derive the radii of the stars, using the Stefan-Boltzman equation and the above numbers for luminosities and temperatures.  We find $R_{\rm O star} = 6.47^{+0.99}_{-0.82} R_\odot$, and $R_{\rm WR}=2.47^{+1.17}_{-1.07} R_\odot$.  The radius uncertainties were derived in the same manner as above, by varying the total absolute visual magnitude (set by the extinction uncertainty), the resulting luminosities and temperatures converted to radii; the minimum and maximum radii obtained define the quoted bounds.   

We remind the reader that this radius we compute for the WN3 star corresponds to the radius at $\tau$=2/3rds.  Again appealing to the unpublished modeling work previously mentioned, we note that for LMC WN3s this radius is roughly $2\times R_*$, where $R_*$ is the ``core" radius taken to be at the bottom of the wind (e.g., $\tau$=100).  The He\,{\sc ii} $\lambda$4686 line formation region peaks about $20\times R_*$, i.e., $10\times$ what we are taking as the radius here. (It extends to $50-100\times R_*$, beyond the orbit of the O star, as discussed later.)  By contrast, the N\,{\sc v} $\lambda \lambda$4604,20 doublet formation region peaks at roughly $2.5\times R_*$, similar to the radius where $\tau$=2/3rds.

In order to measure the projected rotational velocity $v\sin{i}$ of the O star, we used the {\sc cmfgen} model mentioned above, convolving it with the instrumental resolution of the Magellan spectra, and then broadening it further with various values for the rotation.  We found the best match with a $v\sin{i}$ of 175 km s$^{-1}$.  The synchronous rotation of the O star with the orbital period would require a rotational speed of $93^{+14}_{-12}$ km~s$^{-1}$, given its $6.47^{+0.99}_{-0.82}R_\odot$ radius and the 3.52~day orbital period.  Thus, the O star is spinning {\it much} faster than synchronous motion with its orbit, regardless of its inclination.

We summarize these results in Table~\ref{tab:resultsA}.

{\it Note:} We were intrigued to see if we could do any better in determining the physical properties of these stars by applying disentangling techniques.  We wrote and implemented a simple ``shift and add" algorithm. (For a clear explanation of the method , see Section 4.3 of \citealt{2019A&A...627A.151S}.)
However, given the large flux ratio, the limited number of spectra, and the good but not spectacular signal-to-noise, we did not succeed in obtaining further information about the WN3 (which might have permitted modeling), or a sufficiently clean spectrum of the O star suitable for quantitative analysis.  The hope had been to obtain a spectrum of the O star without the contamination of faint He\,{\sc ii} emission features from the WR.  It turned out that we did better by simply using the spectra for which the radial velocities did that for us.

\begin{deluxetable}{l c  c}
%\tabletypesize{small}
\tablecaption{\label{tab:resultsA} Physical Parameters Derived from the Spectra}
\tablehead{
\colhead{Parameter}
&\colhead{O Star}
&\colhead{WR Star}
}
\startdata
$T_{\rm eff}$& 40,000$\pm$500 & 61,000$\pm$12,000 \\
$M_V$ & $-4.01^{+0.28}_{-0.30}$ & $-2.50^{+0.42}_{-0.39}$ \\
$\log{L/L_\odot}$ & $4.99\pm0.07$ & $4.88^{+0.39}_{-0.58}$ \\
$R/R_\odot$ & $6.47^{+0.99}_{-0.82}$ & $2.47^{+1.17}_{-1.07}$ \\
$v\sin{i}$ (km s$^{-1}$)& 175& \nodata\\
Sync.\ rot.\ (km s$^{-1}$)& $93^{+14}_{-12}$  & $35^{+17}_{-15}$ \\
\enddata
\end{deluxetable}

\subsection{Radial Velocity Measurements and the Mass Ratio}
In order to measure the radial velocities of the two components, we used {\sc splot} within {\sc iraf} to measure the wavelengths of the least contaminated absorption lines as well as the N\,{\sc v} $\lambda \lambda 4604,20$ emission doublet. Examples of two of our spectra are presented in Figure~\ref{fig:spect}.

Initial calculations of the radial velocities of the O star included measurements from the upper Balmer lines from H8-H12, H$\delta$, He\,{\sc ii} $\lambda$4200 H$\gamma$, He\,{\sc i} $\lambda$4471, and He\,{\sc ii} $\lambda$4542. However, we found that the H10-12 lines were often discordant relative to the other lines over all of the observations. This was due both to the poorer S/N in the far blue and the scarcity of comparison lines, and thus we excluded those lines from further analysis. The H$\beta$ line has a significant contamination with He\,{\sc ii} emission from the WR, and so we did not use that line either. That left us with seven good lines for our calculations of the radial velocity of the O star.  We were pleased to discover that despite the obvious contamination of the EWs in the He\,{\sc ii} $\lambda$4200 and $\lambda$4542 lines, their velocities agreed well with the stronger H$\delta$, H$\gamma$, and the uncontaminated He\,{\sc i} $\lambda$4471 line.  The He\,{\sc i} $\lambda$5876 line was not used because of the contamination by the Galactic Na~D interstellar lines.

For the absorption spectrum, each line was measured using a singular Gaussian fit. A radial velocity was calculated for each line using the Doppler formula. The means, standard deviations, and medians were  calculated for each spectrum. The medians are presented in Table 1, after applying the heliocentric correction. The standard deviation averaged over each observation date was 6.9 km s$^{-1}$. 

As for the WN3 star, only the NV $\lambda\lambda$4604,20 doublet was selected for analysis, as it was the strongest, cleanest line in the spectrum.\footnote{It is well known from other studies (c.f., \citealt{1944ApJ....99..273H, 1973PASP...85..220N, 1981ApJ...244..157M,2018A&A...616A.103S}) that the He\,{\sc ii} $\lambda$4686 is a poor measure of the motion of the underlying star as it forms in a very extended region, and is easily perturbed by interactions from a companion, as well as wind variability \citep{2019A&A...627A.151S}. By contrast, the N\,{\sc v} doublet is formed close to the core of the star, as previously mentioned.} This was measured using a centroid, five times for each
spectrum.  We then calculated the corresponding radial velocities, and the average of those was selected for the WR radial velocity in Table 1 after applying the heliocentric correction. Uncertainties were determined from the repeatability of the measurements. We estimate this uncertainty as 41 km s$^{-1}$, not unexpectedly large given that the line has a width (base-to-base) of $\sim 4900$ km s$^{-1}$.

\begin{figure}
% \epsscale{0.9}
 \plotone{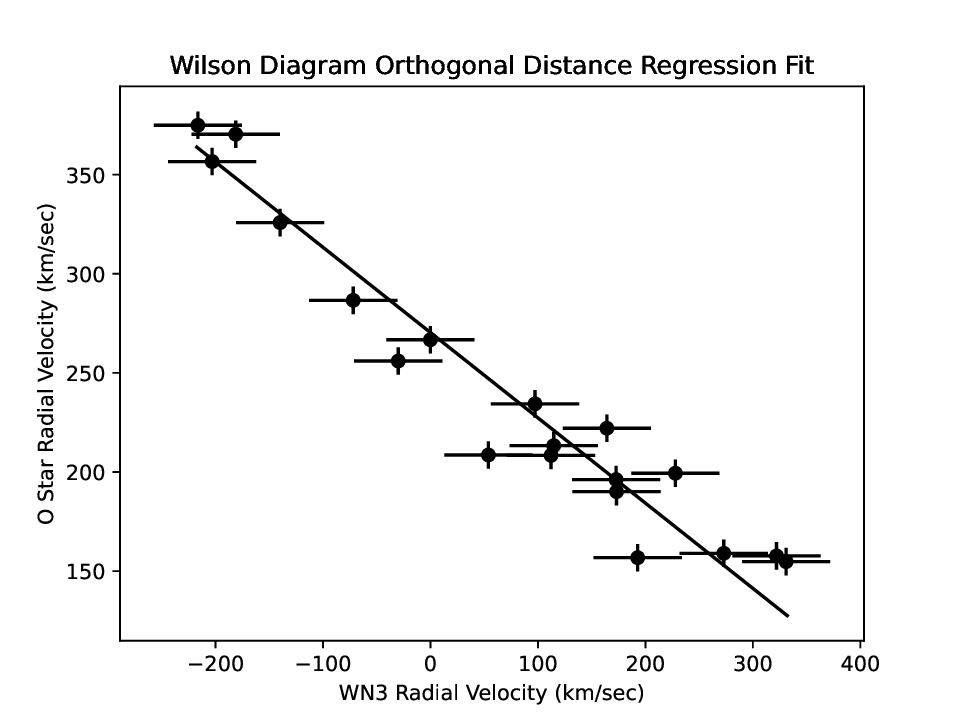}
 \caption{\label{fig:wilson}Radial velocity comparison of the O and WN3 stars. The data points are plotted showing the typical velocity uncertainties for each component, 6.9 km s$^{-1}$ for the O star, and 41 km s$^{-1}$ for the WN3 component. The line shows the best fit for the data points, taking their uncertainties into account.}
 \end{figure}

\citet{1941ApJ....93...29W} called attention to a straight-forward way to measure the mass ratio of a double-lined spectroscopic binary.  If one plots the radial velocity of one star against that
of the other, the relationship should be a straight line whose slope is the ratio of the masses of the two components.  This is a powerful tool, as the result is independent of any of the orbital parameters, and thus provides a check on the ratio of the orbital semi-amplitudes derived from any orbit. With the calculated radial velocities, we were able to determine the mass ratio between the WN3 and the O star via such a Wilson diagram, as shown in Figure~\ref{fig:wilson}. The method for constructing this diagram includes calculating the orthogonal distance regression, taking  the uncertainties in the radial velocities into account. The slope of the regression line is -0.430 $\pm$ 0.028. The $R^2$ value (coefficient of determination) is 0.927. 

This result is what initially intrigued us about this system, as it shows that the mass of the WR star is only 43\% of that of the O star.  Yet, the fact that it is the more evolved object means that it must have lost a significant amount of its mass. If all of the mass lost by the WN star progenitor was accreted by the companion, then the WN progenitor must have lost at least $\sim$30\% of its original mass. If none of the mass were accreted by the companion, then it has lost at least $\sim$60\%. We note here that at the low metallicity of the LMC (1/2 solar; see discussion and references in Section 3.4 of \citealt{RSGWRs}) this simply could not happen by stellar winds alone, but strongly implies that binary evolution has been at play.  We explore this further in Section~\ref{Sec-discussion}.

\subsection{The Orbit}
\label{Sec-orbit}

With the addition of the Gemini spectra, we found we could readily find the value of the period.  We ran the heliocentric radial velocities from Table~\ref{tab:RVs} through a Lafler-Kinman period search program \citep{1965ApJS...11..216L}.  The Lafler-Kinman method is similar to the phase dispersion minimization (PDM) method, but does not utilize bins.  Instead, for each trial period, the data are sorted by phase, and the point-to-point scatter is computed, and compared to the overall scatter of the entire data set. The next trial period is selected making sure that the step size is small enough to maintain good phase integrity across the entire data set.  Our experience is that this works better than PDM when the number of data points is small in number (say, 10) rather than large (many dozen). 

As previously noted, a casual inspection of the data suggested that the period was on the order of several days, given the large radial velocity change between 2021 Jan 22 and 23.  We were also confident that the lack of radial velocity change between 2025 Feb 6 and 7 would constrain possible values.

The Lafler-Kinman search revealed a strong minimum at about 3.5210$\pm$0.0002 days for both the emission and absorption line velocities.  We adopted this as our preliminary value, and used a differential corrections program based on \citet{1967mamt.book..251W} to compute preliminary values for the other elements.  This confirmed our expectations that the orbit was purely circular: given the short period and high masses, tidal forces are expected to quickly circularize the orbit.  For luminous stars with strong winds, we expect that the two components will have different ``center of mass" $\gamma$ velocities as even the photospheres are affected by winds. This is particularly true for WR binaries where all the emission is formed in outflowing winds (see, e.g., \citealt{MasseyA1} and references therein).  In the case here, there is also significant uncertainty in what effective wavelength to use for the unresolved N\,{\sc v} $\lambda \lambda 4604,20$ doublet.  Therefore, we allowed the $\gamma$ velocities to be an independent parameter for each star, along with the orbital semi-amplitudes $K$.

\begin{figure}
\plotone{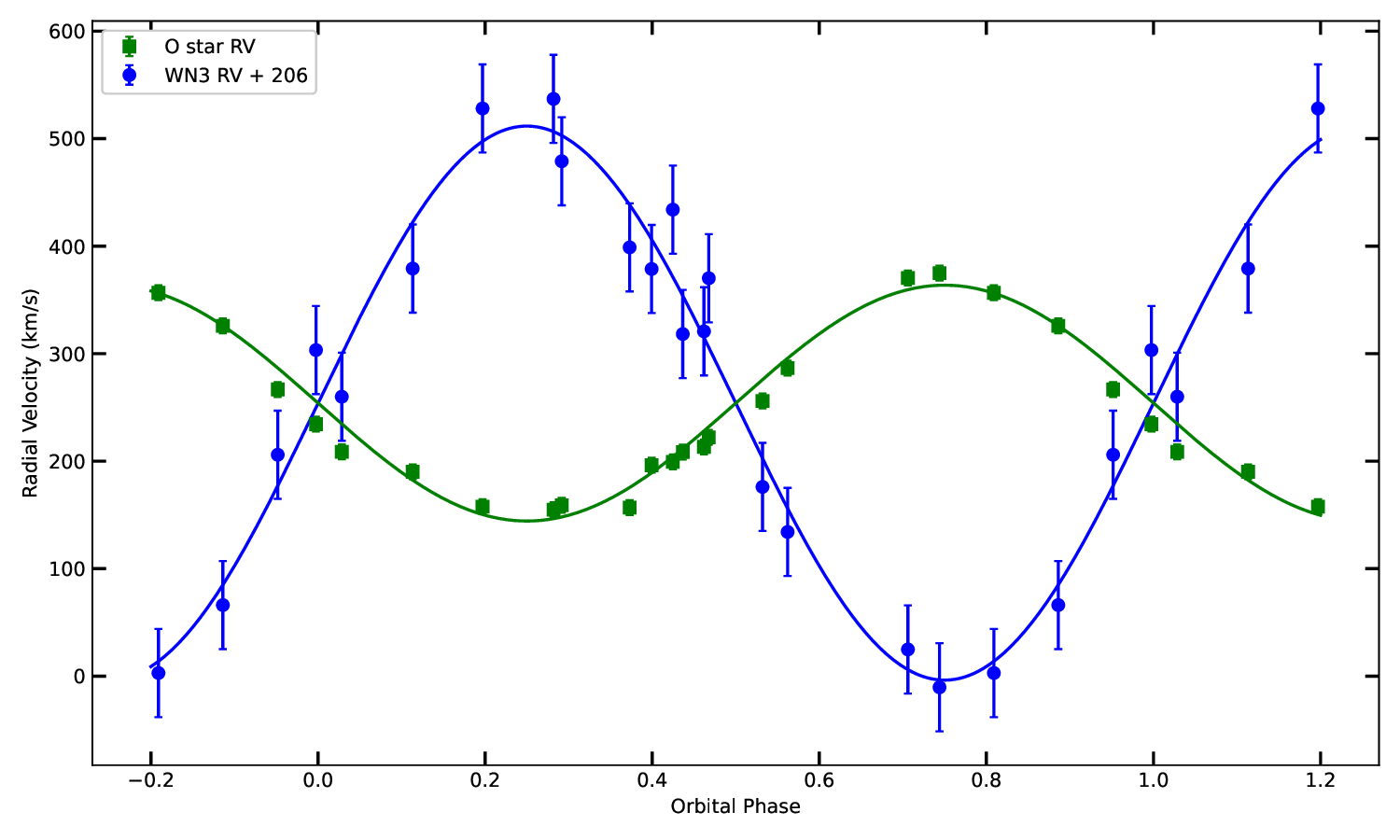}
\caption{\label{fig:orbit} Orbit solutions for LMC173-1.  The radial velocities of the O star are shown as green squares, and its radial velocity curve by a green solid line.  The radial velocities of the WN3 star are shown by blue circles, and the corresponding radial velocity curve by a blue line. We have included the error bars of the typical velocity uncertainties, 6.9 km s$^{-1}$ for the O star primary, and 41 km s$^{-1}$ for the WN3 secondary.  We have offset the radial velocities of the WR star by 206 km s$^{-1}$ in order to match their $\gamma$-velocities for the purposes of illustration. The data from phases 0.8-1.0 are repeated as -0.2 to 0.0, and the data from phases 0.0-0.2 are repeated as 1.0-1.2.}
\end{figure}

For our final solution, we needed to fix both the period and the phase zero-point $T0$ to be the same for the two components.  We therefore computed a least-squares solution allowing $\gamma_{\rm WN3}$, $\gamma_{\rm O star}$, $K_{\rm WN3}$, $K_{\rm O star}$ to vary.  By utilizing the radial velocity information from Table~\ref{tab:RVs} for both stars, we forced $P$ and $T0$ to be the same for the components.  We have adopted the phase zero-point (corresponding to time $T0$) to be when the WR star is in inferior conjunction.  This is consistent with the convention that the brighter and more massive star is being eclipsed at phase 0.   The eccentricity was set to zero. All data were weighted the same. We note that the residuals from the orbit (RMS) are comparable to the measuring uncertainties discussed in the previous section, namely 41 km s$^{-1}$ for the WN3 star, and 6.9 km s$^{-1}$ for the O star.  Our results are given in Table~\ref{tab:orbit}, and shown in Figure~\ref{fig:orbit}.

We see that the derived mass ratio, $0.426\pm0.044$, agrees very well with the mass ratio derived from the Wilson diagram in the previous section, $0.430 \pm 0.028$. As emphasized earlier, the latter value does not depend upon the values of the orbital eccentricity or period.   The consistency of these values is in accord with our assumptions.

For the derived quantities (the minimum masses and the projected orbital separations) we used a Monte Carlo sampling of the periods and orbital semi-amplitudes assuming Gaussian distributions with sigma values corresponding to the uncertainties, following the recent analysis of \citet{MasseyA1}. This allows us to sample the full posterior distributions of the inputs, thus capturing the true asymmetric and skewed uncertainties of the derived quantities.  This is important because the mass function depends linearly on the period but cubically on the velocity amplitudes, making the use of linear approximation inadequate in general.  In practice, the minimum masses calculated in this way agree with the values computed more straightforwardly, to the quoted precision.

\begin{deluxetable}{l c  c}
%\tabletypesize{small}
\tablecaption{\label{tab:orbit} Orbital Fits and Results for LMC 173-1}
\tablehead{
\colhead{Parameter}
&\colhead{O Star}
&\colhead{WN3}
}
\startdata
$P$ (days) & \multicolumn{2}{c}{$3.521120\pm0.000048$} \\
$T0$ (HJD) & \multicolumn{2}{c}{$2456588.91\pm0.04$} \\
$e$ (fixed) &\multicolumn{2}{c}{0.00} \\
$\gamma$ (km s$^{-1}$)& $254.0\pm6.7$ & $48.0\pm6.8$  \\
$K$ (km s$^{-1}$)   & $109.7\pm10.3$  & $257.7\pm10.5$ \\
Mass ratio (WR/O Star)& \multicolumn{2}{c}{$0.426\pm0.044$} \\
RMS (km s$^{-1}$)   & 13.2  & 35.0 \\
%$a\sin{i}$ ($10^6$ km) & $12.5\pm0.5$& $5.3\pm0.5$  \\
$a\sin{i}$ ($R_\odot$)& $7.6\pm0.7$  & $18.0\pm0.7$\\
$a\sin{i}$(total)($R_\odot$) & \multicolumn{2}{c}{$25.6\pm1.4$} \\
$m\sin^3{i}$ ($M_\odot$)  & $12.7\pm1.4$ & $5.4\pm0.9$ \\
Orbital inclination  (hard limit) $i$ & \multicolumn{2}{c}{$\lesssim71^\circ$} \\
Orbital inclination (modeling) $i$ & \multicolumn{2}{c}{$50^\circ\pm7^\circ$}\\
masses $m (M_\odot$) $i<71^\circ$ & $>$$15.0\pm1.7$ & $>$$6.4\pm1.1$\\
masses $m (M_\odot$) $i=50^\circ\pm7^\circ$  & 2$8.2^{+6.9}_{-5.1}$& $12.0^{+3.3}_{-2.6}$\\
\enddata
\end{deluxetable} 

\section{The Light-curve and Orbital Inclination}
\label{Sec-lc}

\subsection{OGLE Photometry}

LMC173-1  has not been cataloged as a variable in any photometric data bases.  The only publicly available time-series are the LMC photometric maps from the OGLE-III shallow survey \citep{2012AcA....62..247U} available via Vizier at  ``J/AcA/62/247/stars."  These show that the star's $V$ and $I$ photometry have sigmas of only 0.012~mag and 0.013~mag, respectively, from 56 and 58 visits, suggesting no obvious variability.  The number of observations made it highly unlikely we missed a significant eclipse that would help determine the orbital inclination; however, since the HJDs of the observations were not available,  we could not be completely certain that an
eclipse had not slipped through the cracks.  
 
\begin{figure}
\plotone{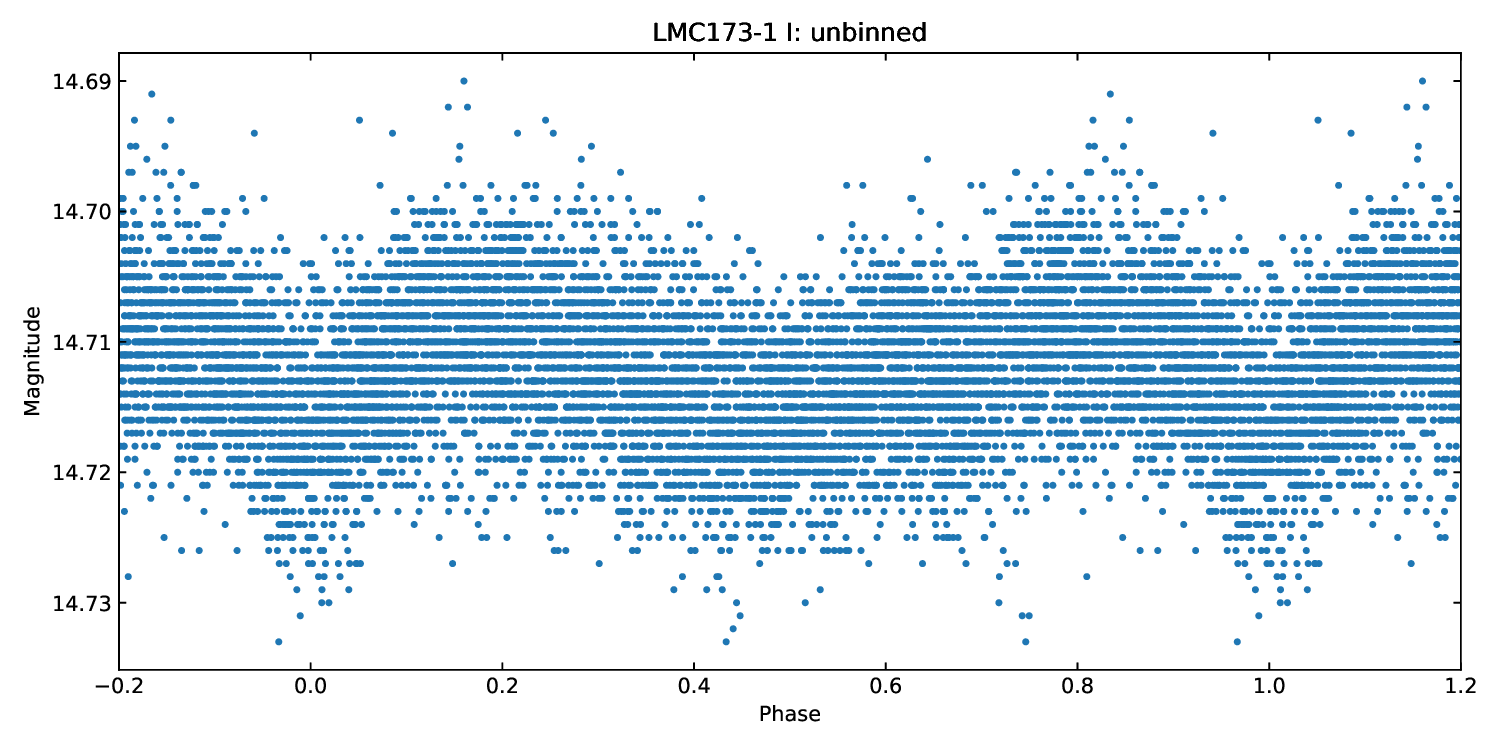}
\plotone{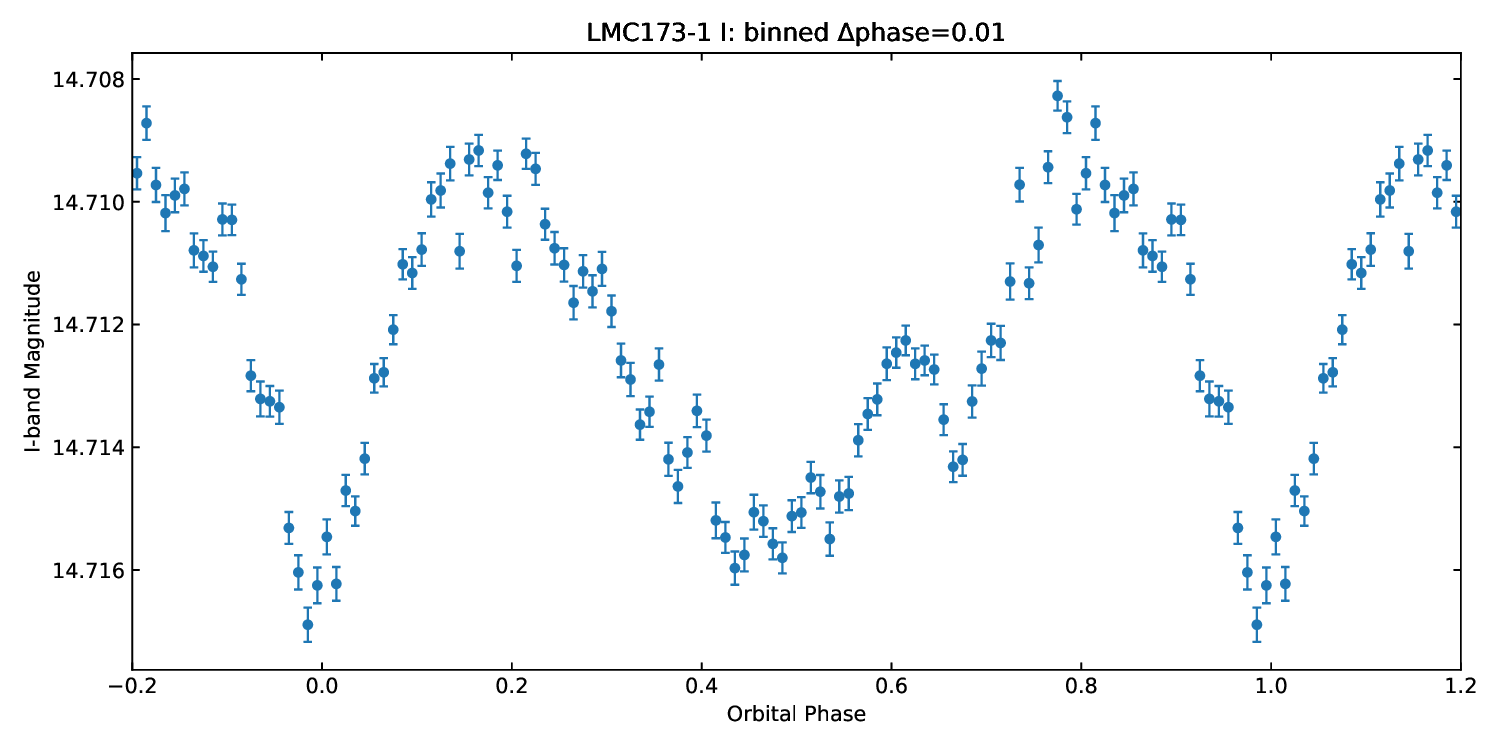}
\caption{\label{fig:Iband} Phased I-band photometry of LMC173-1.  The upper figure shows the unbinned data; no error bars are shown to avoid crowding, but are typically 2-3 millimag.  The lower figure shows the data binned by 0.01 phases; the error bars show the resulting standard deviations of the mean within each bin, typically 0.2-0.3 millimag.}
\end{figure}

N.I.M contacted the OGLE team for further information, with the result that M.K.S. and A.U. kindly provided us with new data that is part of their on-going OGLE-IV survey. These data are of higher photometric precision, and much higher cadence than those of OGLE-III. The I-band data contains 7961 observations.  The data begin in March 2010, but the vast majority have been acquired during the three full Magellanic Cloud observing seasons in 2022/2023 through 2024/2025, plus the partial 2025/2026 season currently underway.  These post-Covid observations consist of sequences of ten to eleven measurements per night, with a regular intra-night cadence of approximately 20 minutes. This dense sampling provides excellent phase coverage and sensitivity to millimag-level variability.  The typical uncertainty of a single observation is 2-3 millimag.  We provide these data here in Table~\ref{tab:Iband}, and show the unbinned data in the upper panel of Figure~\ref{fig:Iband}.

\begin{deluxetable}{r r r r}
\tablecaption{\label{tab:Iband} OGLE-IV I-band Photometry}
\tablehead{
\colhead{HJD-2450000}
&\colhead{Phase\tablenotemark{a}}
&\colhead{Mag}
&\colhead{Uncert.}
}
\startdata
 5260.64086 & 0.77084 & 14.702 &  0.003 \\
 5261.64422 & 0.05579 & 14.718 &  0.003 \\
 5262.63801 & 0.33803 & 14.713 &  0.003 \\
 5264.60542 & 0.89678 & 14.700 &  0.003 \\
 5265.60667 & 0.18113 & 14.706 &  0.003 \\
  \enddata
 \tablecomments{Table~\ref{tab:Iband} is published in its entirety in the electronic 
edition of the {\it Astrophysical Journal}.  A portion is shown here 
for guidance regarding its form and content.}
\tablenotetext{a}{Computed using P=3.521120 d and phase zero-point T0=2456588.91.}
\end{deluxetable}

When phased to our spectroscopic ephemeris, the new data confirm that there are no actual geometric eclipses present.  However, the data do reveal some interesting structure.  Given the dense coverage in phase space, we were able to smooth the data by 
0.01 phase bins in order to better reveal these features.  These binned data have precision of 0.2 - 0.3 millimag, allowing the detection of low-level structure. The binned data are shown in the lower panel of Figure~\ref{fig:Iband}.

We see that the light-curve is dominated by a double-wave ellipsoidal modulation, with a peak-to-peak amplitude of 7-8 millimag.  This variation must be dominated by the O star, as it provides the majority of the optical flux.  Furthermore it has a radius several times that of the WR and must be much closer to filling its Roche lobe, as the tidal distortion scales very steeply with stellar radius $R$, i.e., as $(R/a)^3$, where $a$ is the separation. \citep{Kipp}.  

In addition to this ellipsoidal modulation, there is a sharper, narrower feature near phase 0 (when the WR star is in front), and a millimag sharp feature around a phase of 0.65. The feature at phase 0 cannot be a grazing geometric eclipse, as an eclipse this small in amplitude would require an exquisitely well aligned geometry.  Rather, this is the classical signature of a WR ``atmospheric eclipse."   This interpretation has a long history in the prototypical eclipsing WR+O system V444 Cyg (e.g., \citealt{1968ApJ...152...89K, 1978ApJ...221..193H,  1984ApJ...281..774C}) and has been discussed in later work incorporating wind and wind–wind collision structure (e.g., \citealt{2002A&A...388..957K, 2012MNRAS.420..495A}). 

We were initially puzzled by the small dip around phase 0.65.  We have confirmed that this feature is seen in each densely observed season, and we conclude that it is associated with the colliding-wind interaction region crossing the line of sight. 

In interpreting the light-curve, it is useful to keep in mind that the O star is actually orbiting within the formation region of the He\,{\sc ii} $\lambda$ 4686 line.  From our WN3 models we find that this region extends to $50-100R_*$, or $25-50R_{\tau=2/3}$. This would be  of order $100R_\odot$, compared to the $25R_\odot$ separation of the two stars!  

There is a much more limited data set available in V.  When phased with our ephemeris it shows the same double-wave modulation seen in I. The sparse sampling does not allow for a detailed comparison of the low-amplitude features. However, we do extract the V magnitude from these data, 14.65, as stated earlier.

\subsection{Orbital Inclination}

Because the I-band variability is dominated by ellipsoidal modulation and wind-related attenuation rather than well-defined geometric eclipses, there are significant challenges to using the light curve to determine the inclination.  In this section we will begin by placing a hard limit on the inclination, and then proceed to model the tidal modulation of the light curve to get an approximate value for the inclination. 

Assuming spherical stars,\footnote{Mooo.} the lack of eclipses requires that  $a\cos{i} > R_{\rm O star} + R_{\rm WN}$, where $a$ is the separation between the two
stars.    Although we do not know $a$, we do know $a\sin{i}$ from the orbit solution (Table~\ref{tab:orbit}), and thus $$i_{\rm max} = \arctan{\frac{a\sin{i}}{R_{\rm O star} + R_{\rm WN}}}.$$ We compute $i\lesssim71^\circ$.

\subsection{Roche-lobe Filling Factors}

We can use the \citet{1983ApJ...268..368E} equation to compute the Roche-lobe radius ($R_L$) for each star:
$$\frac{R_L}{a} = \frac{0.49 q^{2/3}} {0.6 q^{2/3} + ln (1+q^{1/3})},$$ where $q$ is the mass ratio of the star whose Roche-lobe radius
you want relative the mass of its companion.  Given that $q_{\rm WN}=0.426$, $q_{\rm O star} = 2.35$,  then $R_L (\rm O\ star)/a = 0.454$ 
and $R_L (\rm WN)/a = 0.308$. 

Using our orbital separations, we have $$R_L (\rm O\ star) \sin{i} = 11.6 R_\odot$$ and $$R_L (\rm WN) \sin{i} = 7.9 R_\odot.$$ We can now compare these
to our stellar radii: $$R_{\rm O star} / R_L ({\rm O\ star}) = 0.56 \sin{i},$$ and $$R_{\rm WN} / R_L ({\rm WN}) = 0.31 \sin{i}.$$  So, even at
$i=90^\circ$ the O star is at 56\% of its Roche-lobe radius and the WN star is at 31\% of its Roche-lobe radius.  Given that $i\lesssim71^\circ$ we can refine these numbers
slightly: the O star's radius is no more than 53\% of its Roche-lobe radius, and the WN star's radius is no more than 30\% of its Roche-lobe radius.  Of course, comparing
the WN's ``radius" to its Roche-lobe radius is a bit slippery since the wind is so extended.  The Roche-lobe limit applies cleanly to the hydrostatic star; the wind can
extend well beyond the WN3's Roche-lobe radius without implying overflow.

\subsection{Modeling the Light Curve}

We can, however, go further than this.  The ellipsoidal modulation of the light curve is tiny, only 7-8 millimag, and is caused primarily by the varying viewing geometry of the tidally distorted O star, as well as
limb/gravity darkening and the reflection effect of the incident flux on the O star.  The situation is complicated by the presence of electron scattering of the O star's light by the inner part of the WR's wind at phase 0 (where the O star is behind) and by the outer part of the wind at phase 0.5 (where the O star is in front but still within the WR star's wind.  We allow for the effect of electron scattering by the wind by
lowering the albedos, as described below. 

Model light curves were constructed adopting our orbital parameters, combined $M_V$ of the system, visual flux ratio, effective temperatures, and radii. We use the light curve synthesis code {\sc gensyn}\citep{1972MNRAS.156...51M}. We obtained V and I fluxes (chosen at the central wavelengths, 5470 \AA\ and  8050 \AA) and limb darkening coefficients from values from \citet{2007ApJS..169...83L} and \citet{1985A&AS...60..471W}, respectively. With the significant difference in the effective temperatures of the companions, we expect that reflection effects will contribute to the photometric variability. For both stars, electron-scattering decreases the amount of incident flux that will be absorbed, effectively lowering the albedos. We adopt albedo values of 0.30 and 0.00 for our O star and WN3, respectively \citep{1971ApJ...167..137H}. 

We were able to construct acceptable model light curves with a large range of inclinations (43$^\circ$ - 57$^\circ$). For all the model fits, the derived radii matched the values in Table 2.  The flux ratios from the modeling
come out as 0.221, compared to our calculation of 0.25$\pm0.15$. The $M_V$ of the system comes out
as $-4.27$, consistent with our calculation of $M_V =-4.24\pm0.16$.
The best fit model has Roche-lobe filling factors of 0.493 and 0.393, for the O-star and WR, respectively, compared to the values derived above, 0.43 and 0.24 for 50$^\circ$. These go up to 0.535 and 0.423 for $57^\circ$, and down to 0.444 and 0.358 for $43^\circ$. 

\begin{figure}
\plotone{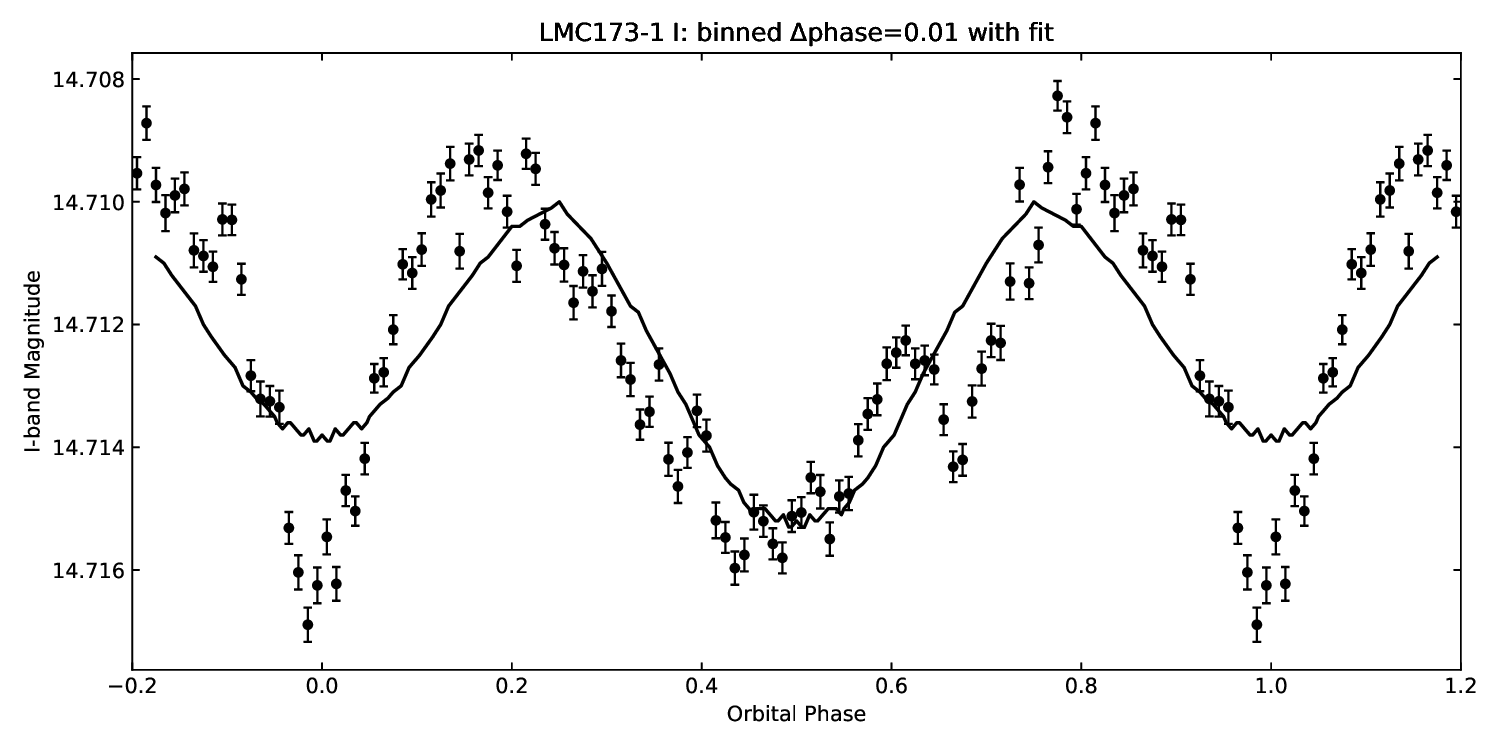}
\caption{\label{fig:Laurafit} Light curve fit.  We have modeled the ellipsoidal I-band light-curve as described in the text.  We found acceptable inclinations of 43$^\circ$ to 57$^\circ$, with the best fit at 50$^\circ$, which we show by the solid line.  The sharp features not modeled (at phase 0 and phase 0.65) are explained as effects from the wind,
as described in the previous section.}
\end{figure}

 Our best-fit model is shown in Figure~\ref{fig:Laurafit}.  We were unable, in any model, to completely match the features near phases 0 and 0.65, as expected.  The slight discrepancies between the model fit and the
 data are not surprising, as they are on the order of a small fraction of a millimag.  The fit depends not only
 on the geometry of the tidal distortion but also on limb darkening/gravity and reflection effects, none of
 which can be taken perfectly into account.   A simple two-harmonic fit 
to the unbinned I-band light curve, completely independent of the detailed 
physical light-curve modeling, recovers a dominant 2f double-wave modulation 
with a peak-to-peak amplitude of about 8 millimag, an excellent match to that of the light-curve
model.
 
We include the derived masses in Table~\ref{tab:orbit}, where the uncertainties were estimated via Monte Carlo simulations drawing the minimum masses from a correlated bivariate normal consistent with the measured mass ratio and sampling the inclination from a truncated Gaussian distribution.

\section{Summary and Discussion: the Evolution of this Massive Binary}
\label{Sec-discussion}

Trying to understand how an evolved binary got to be the way it is today is a little like trying to un-spill milk.  Let us begin by summarizing what we know about this interesting system, and then describing a plausible scenario.  We will then see if modern binary
models will allow us to put some of the milk back in the bottle.

Currently, the WN3 star has a mass that is 43\% as much as its O star companion.  The two stars are in a 3.52 day circular orbit about their common center of mass. Unfortunately, the viewing angle is such that there are no geometric eclipses, placing only an upper limit on the orbital inclination and hence a lower limit on the actual masses.  The O  star is rotating faster than it would in synchronous rotation with the orbit, but not as fast as many other spun-up companions (see, e.g., \citealt{MasseyA1}).  Neither star is close to filling its Roche-lobe surface.
Nevertheless, at one time the progenitor of the WN3 must have transferred a great deal of mass to its companion, as it is the more evolved object and therefore must have evolved faster.

Throughout this paper we have insisted that the WN3 could only have formed via binary evolution.   This conclusion is based on the hypothesis that stellar-wind mass loss could not have resulted in the more evolved star having a mass only 43\% as much as its companion.  Let us examine the possibility that single-star evolution is sufficient to explain this system.   The progenitor of the WN3 must have begun
its life as the more massive member of the system, so let us understand what the current spectral type of the O star companion implies.
According to the single-star Geneva evolutionary models computed for LMC-like metallicity, \citep{2021A&A...652A.137E} only stars with
initial masses of $32M_\odot$ ever become hotter than 40,000~K on the main-sequence. (Of course, all bets are off if the O5.5~V star
had accreted mass, but here we are considering only a single-star scenario.) A likely match would be a 3.5-4 Myr 40$M_\odot$ at $Z=0.006$, similar to what we would find at Galactic metallicity from the \citet{Sylvia} models (see, e.g., Table 
1 in \citealt{2017RSPTA.37560267M}).  Higher mass stars never become that cool while on the main-sequence (Figure~1b in \citealt{2021A&A...652A.137E}).  Thus whatever mass the WN3 progenitor started with (we only know more than that of the O star), it has managed to lose enough mass to become a $17M_\odot$ ($0.43 \times$ the mass of the O star), given that the O companion's mass has not changed much during this time.  Is this much mass loss reasonable given the metallicity of the LMC?

\citet{2021A&A...652A.137E} adapts the standard mass-loss formulation of \citet{2001A&A...369..574V} for the metallicity of the LMC. According to their models, if the progenitor of the WN3 star had started with a mass of 60$M_\odot$, by the time it was at its bluest part of He-burning, its mass would have only decreased to 36$M_\odot$.  In any event, single stars of 60$M_\odot$ are not expected to become WRs at the metallicity of the LMC.  According to the Geneva models, only the highest mass stars (e.g., $85M_\odot$ and $120M_\odot$ in their grid) are expected to become WNs.
At 4.0~Myr, an $85M_\odot$ will have the prerequisite characteristics to be considered a WN star; however, its mass will be $35M_\odot$, not 17$M_\odot$, even at the end of carbon burning.  According to the models, even a $120M_\odot$ ends its life at $52M_\odot$.   Of course, these models may not fully take into account all possible sources of mass loss.  The role of mass loss during the enigmatic 
Luminous Blue Variable (LBV) stage is not explicitly included, as this requires treatment of the ``atmospheric Eddington Limit," (e.g., 
\citealt{2025ApJ...981..176L} and references therein), as line opacities are not accounted for in these interior evolutionary calculations.  

However, even if LBV episodic mass loss stripped off additional material, there is a further problem with the single-star scenario: the luminosity of the O star companion itself.  We measure 
 $\log L/L_\odot=4.99\pm0.07$, but this low a luminosity does not correspond to a $40M_\odot$ star.  Rather that
 luminosity is characteristic of a main-sequence $25M_\odot$ star, which will never have an effective temperature as hot as 40,000~K.  The luminosity is consistent with the mass we measure (see Table 3), but not the temperature---unless the O star is the product of binary ``rejuvenation."  \footnote{We take this opportunity to note that the WN3 star's luminosity $\log L/L_\odot=4.9^{+0.4}_{-0.6}$ is nearly a factor of 2 lower than the minimum luminosity found by \citet{2020A&A...634A..79S} for WR stars in the LMC, although we do not consider the LMC WR luminosity function or cutoff to be well established yet.}
 
 Finally, we present the least model-dependent argument: the current period of the system is 3.52~d.  If
 the WN3 progenitor lost its mass via stellar winds, and this mass was lost to the system, the initial period
 of the system must have been much shorter.  Back evolving the orbit under Jeans-mode wind loss (\citealt{1924MNRAS..85....2J}; see also \citealt{1971ARA&A...9..183P}) we calculate that the original orbital period would have been 1.7-1.8 d.  In that case the Roche-lobe radii for the star would be 9$R_\odot$, comparable to the early-time radius for a 30$M_\odot$ star, and RLOF would have been unavoidable, invalidating the assumption that isotropic stellar winds were responsible. 
 
Instead, the evolution of this star is more easily explained by the following scenario.
The initially more massive star filled its Roche-lobe surface before leaving the main sequence, corresponding to Case A Roche-lobe overflow (RLOF).
As the donor star lost most of its hydrogen-rich envelope, it contracted dramatically, detaching from its Roche-lobe surface and emerging as a WN-type Wolf–Rayet star. The companion was rejuvenated by the accretion, and the orbit first shrank as the mass ratio approached unity and then re-expanded after mass-ratio reversal. Subsequent mass loss due to the WN star’s strong stellar wind has likely removed some additional mass and angular momentum from the system, further widening the orbit to what we see today.  The resulting system, a detached, short-period WN + O binary with nearly equal luminosities but very different radii, thus represents the post-RLOF phase following mass-ratio reversal and orbital re-expansion.

A search of the extensive BPASS v2.2 binary evolutionary models \citep{BPASS2} 
reveals a narrow
range of possible progenitors that could lead to the current
mass ratio, period, and luminosities.  In the best-matched model,
the progenitors had initial masses of 22.3$\pm$1.0 $M_\odot$ and
16.3$\pm$0.5 $M_\odot$, with an initial orbital period shorter than
now, 2.5 d.  The model's current masses are 12.3$\pm$0.3 $M_\odot$ 
and 25.9$\pm$0.7 $M_\odot$, suggesting an orbital inclination of
about 50$^\circ$. This is in remarkably good agreement with the inclination we derived from
modeling the light-curve. The age would be 8.5~Myr. In this model, mass transfer was highly efficient, with only 0.4$M_\odot$ lost to the system.  With the companion doubling in mass, one expects a significant spin up.  An inclination of 50$^\circ$ yields an equatorial rotation speed of 230 km~s$^{-1}$ (based upon our observed $v\sin{i}$) which, while fast, is not phenomenal, suggesting that the mass transfer took place over a nuclear timescale, and not a thermal timescale, allowing tidal forces to limit the spinup.

One of the outstanding problems in massive star evolution is the relative importance of binary evolution versus stellar winds in terms of the formation of WR stars.  Over the years studies have consistently shown that the close binary frequency of WR stars is 30-40\%, independent of environment \citep{2003MNRAS.338..360F,2003MNRAS.338.1025F,2008MNRAS.389..806S,NeugentBinaries}, but binary enthusiasts argue that the rest may have undergone mergers.  We suspect that no single explanation is responsible for the origin of {\it all} WRs. As \citet{BPASS2} put it, ``Single-star stellar winds are not strong enough to create every WR star we see in the sky."  At the same time, binary evolution is unlikely to be responsible for all WRs either.   Given the inequality in the mass ratio in the LMC173-1 system, we conclude that binary evolution is clearly responsible for the formation of its WN3 component.

\begin{acknowledgements}

Lowell Observatory sits at the base of mountains sacred to tribes throughout the region. We honor their past, present, and future generations, who have lived here for millennia and will forever call this place home.  

We are grateful for the excellent support we always receive at Las Campanas Observatory.  The Gemini-S GMOS data were taken by  Drs.\ Rodrigo Carrasco, Emily Deibert, Venu Kalari, Murilo Marinell, and Swayamtrupta Panda, and we thank them for their expert efforts. 

Brian Skiff helped us try to track down archival photometry of LMC173-1.  Dr.\ Deidre Hunter provided much useful advise throughout
the process. We utilized ChatGPT \citep{2023arXiv230308774O} in the construction of  several Jupyter notebooks,
as well as assistance in other areas, such as the Monte Carlo calculations for our error propagations.  A critical review by an anonymous referee prompted us to improve the paper in several regards.

This project was supported in part by the National Science Foundation through AST-2307594, as well as the NAU division of the Arizona Space Grant Consortium, under Federal Grant FAIN: 80NSSC20M0041. L.C.M.'s participation was supported through the Research Experiences for Undergraduates program funded through NAU thanks to the NSF grants 1950901 and 2349774.  The OGLE project has received funding from the Polish National Science
Centre grant OPUS-28 2024/55/B/ST9/00447 awarded to A.U.

NOIRLab IRAF is distributed by the Community Science and Data Center at NSF NOIRLab, which is managed by the AURA under a cooperative agreement with the NSF.

\end{acknowledgements}

\facilities{Magellan:Clay (MagE), Magellan:Baade (MagE), Gemini:South (GMOS), OGLE}
\bibliographystyle{aasjournalv7}
\bibliography{masterbib.bib}

\end{document}